\documentclass{aa}
\usepackage{txfonts}
\usepackage{graphicx}
\usepackage{natbib}
\bibpunct{(}{)}{;}{a}{}{,} 

\begin{document}

\title{Circumstellar material in the \object{Vega} inner system revealed by CHARA/FLUOR}
\titlerunning{Circumstellar material in the Vega inner system}

\author{O.~Absil\inst{1}\fnmsep\thanks{O.~A. acknowledges the financial support of the Belgian
National Fund for Scientific Research (FNRS).} \and E.~di~Folco\inst{2,3} \and A.~M\'erand\inst{2}
\and J.-C.~Augereau \inst{4} \and V.~Coud\'e~du~Foresto\inst{2} \and J.~P.~Aufdenberg\inst{5} \and
P.~Kervella\inst{2} \and S.~T.~Ridgway\inst{5,6} \and D.~H.~Berger\inst{6} \and
T.~A.~ten~Brummelaar\inst{6} \and J.~Sturmann\inst{6} \and L.~Sturmann\inst{6} \and
N.~H.~Turner\inst{6} \and H.~A.~McAlister\inst{6}}

\offprints{O.~Absil}

\institute{Institut d'Astrophysique et de G\'eophysique, Universit\'e de Li\`ege, 17 All\'ee du Six
 Ao\^ut, B-4000 Li\`ege, Belgium \\ \email{absil@astro.ulg.ac.be}
 \and LESIA, UMR8109, Observatoire de Paris-Meudon, 5 place Jules Janssen, F-92195 Meudon, France
 \and Observatoire de Gen\`eve, 51 chemin des Maillettes, CH-1290 Sauverny, Switzerland
 \and Laboratoire d'Astrophysique de l'Observatoire de Grenoble, UMR CNRS/UJF 5571, BP 53, F-38041
 Grenoble Cedex 9, France
 \and National Optical Astronomical Observatory, 950 North Cherry Avenue, Tucson, AZ 85719, USA
 \and Center for High Angular Resolution Astronomy, Georgia State University, PO Box 3969, Atlanta,
      Georgia 30302-3965, USA}

\date{Received 16 November 2005 / Accepted 6 February 2006}

\abstract
 {Only a handful of debris disks have been imaged up to now. Due to the need for high dynamic range
and high angular resolution, very little is known about the inner planetary region, where small
amounts of warm dust are expected to be found.}
 {We investigate the close neighbourhood of Vega with the help of infrared stellar interferometry
and estimate the integrated $K$-band flux originating from the central 8~AU of the debris disk.}
 {We performed precise visibility measurements at both short ($\sim$30~m) and long ($\sim$150~m)
baselines with the FLUOR beam-combiner installed at the CHARA Array (Mt Wilson, California) in
order to separately resolve the emissions from the extended debris disk (short baselines) and from
the stellar photosphere (long baselines).}
 {After revising Vega's $K$-band angular diameter ($\theta_{\rm UD} = 3.202 \pm 0.005$~mas), we
show that a significant deficit in squared visibility ($\Delta V^2 = 1.88 \pm 0.34$\%) is detected
at short baselines with respect to the best-fit uniform disk stellar model. This deficit can be
either attributed to the presence of a low-mass stellar companion around Vega, or as the signature
of the thermal and scattered emissions from the debris disk. We show that the presence of a close
companion is highly unlikely, as well as other possible perturbations (stellar morphology,
calibration), and deduce that we have most probably detected the presence of dust in the close
neighbourhood of Vega. The resulting flux ratio between the stellar photosphere and the debris disk
amounts to $1.29 \pm 0.19$\% within the FLUOR field-of-view ($\sim$7.8~AU). Finally, we complement
our $K$-band study with archival photometric and interferometric data in order to evaluate the main
physical properties of the inner dust disk. The inferred properties suggest that the Vega system
could be currently undergoing major dynamical perturbations.}
 {}


\maketitle


\section{Introduction}

Vega (HD~172167, A0V, 7.76~pc) is probably one of the most important stars in astrophysics, as it
has been used as a photometric standard for more than a century \citep{Hearnshaw96}. However, with
the advent of infrared space-based telescopes, it was discovered to have a large infrared excess
beyond 12~$\mu$m with respect to its expected photospheric flux \citep{Aumann84}. This was
identified as the thermal emission from a circumstellar disk of cool dust located at about 85~AU
from Vega. Since this first discovery of a circumstellar dust around a main-sequence (MS) star,
photometric surveys with IRAS \citep{Fajardo99} and ISO \citep{Laureijs02} have shown that about
10\% of MS stars have significant infrared excess in the $20-25$~$\mu$m region.

\begin{table*}[t]
\caption{Individual measurements. Columns are: (1, 2) date and time of observation; (3, 4)
projected baseline length and position angle (measured East of North); (5) squared visibility after
calibration and error; (6, 9) HD number of calibrators used prior and after the given data point
respectively, 0 means that there was no calibrator; (7, 8, 10, 11) quantities used for computing
the correlation matrix as in Eq.~(26) of \citet{Perrin03}: $\sigma_{V^2}$ are errors on the
estimated visibility of the calibrators.} \label{tab:obs} \centering
\begin{tabular}{ccccccccccc}
\hline\hline & & Projected & Position & Calibrated $V^2$ &  &  &  &  &
\\ Date & UT & baseline (m) & angle ($\degr$) & ($\times 100$) & HD$_a$ & $\alpha$ &
$\sigma_{V^2_a}$ & HD$_b$ & $\beta$ & $\sigma_{V^2_b}$
\\ \hline
  2005/05/21
   & 06:17 & 101.60 &  -76.85 & $20.4 \pm 1.14$ &      0 & 0.000 & 0.000 & 165683 & 0.330 & 0.870
\\ & 07:31 & 127.86 &  -90.04 & $ 6.1 \pm 0.25$ & 176527 & 0.050 & 0.870 & 176527 & 0.060 & 1.025
\\ & 08:20 & 141.07 &  -97.43 & $ 2.6 \pm 0.08$ & 176527 & 0.026 & 1.025 & 173780 & 0.039 & 0.896
\\ & 08:59 & 148.55 & -102.96 & $ 1.3 \pm 0.04$ & 173780 & 0.024 & 0.896 & 173780 & 0.017 & 0.895
\\ 2005/05/22
   & 06:05 &  98.63 &  -75.25 & $23.2 \pm 0.22$ & 159501 & 0.240 & 0.467 & 159501 & 0.064 & 0.624
\\ & 06:24 & 105.77 &  -79.02 & $18.2 \pm 0.20$ & 159501 & 0.142 & 0.467 & 159501 & 0.101 & 0.624
\\ & 06:29 & 107.70 &  -80.00 & $16.8 \pm 0.18$ & 159501 & 0.120 & 0.467 & 159501 & 0.107 & 0.624
\\ & 06:39 & 111.61 &  -81.97 & $14.4 \pm 0.15$ & 159501 & 0.082 & 0.467 & 159501 & 0.115 & 0.624
\\ & 06:49 & 115.39 &  -83.83 & $12.2 \pm 0.14$ & 159501 & 0.052 & 0.467 & 159501 & 0.117 & 0.624
\\ & 06:59 & 118.79 &  -85.51 & $10.4 \pm 0.12$ & 159501 & 0.030 & 0.467 & 159501 & 0.115 & 0.624
\\ & 08:18 & 141.45 &  -97.68 & $ 2.6 \pm 0.07$ & 173780 & 0.014 & 0.624 & 173780 & 0.051 & 0.897
\\ & 08:23 & 142.62 &  -98.45 & $ 2.4 \pm 0.06$ & 173780 & 0.011 & 0.624 & 173780 & 0.051 & 0.897
\\ & 08:34 & 144.75 &  -99.93 & $ 2.0 \pm 0.06$ & 173780 & 0.005 & 0.624 & 173780 & 0.049 & 0.897
\\ 2005/06/13
   & 05:22 &  33.59 &   20.55 & $84.2 \pm 1.42$ & 168775 & 0.543 & 0.152 & 168775 & 0.362 & 0.153
\\ & 06:15 &  33.85 &   13.58 & $83.4 \pm 0.92$ & 168775 & 0.269 & 0.153 & 168775 & 0.628 & 0.153
\\ & 06:46 &  33.92 &    9.33 & $84.5 \pm 0.73$ & 168775 & 0.419 & 0.153 & 163770 & 0.523 & 0.272
\\ & 07:14 &  33.96 &    5.49 & $80.8 \pm 0.99$ & 163770 & 0.510 & 0.272 & 163770 & 0.419 & 0.272
\\ & 07:43 &  33.97 &    1.37 & $82.8 \pm 1.35$ & 163770 & 0.514 & 0.272 & 163770 & 0.438 & 0.272
\\ & 08:13 &  33.97 &   -2.95 & $84.5 \pm 1.19$ & 163770 & 0.833 & 0.272 & 168775 & 0.129 & 0.152
\\ & 09:37 &  33.82 &  -14.54 & $83.6 \pm 0.67$ & 163770 & 0.123 & 0.272 & 168775 & 0.784 & 0.152
\\ & 10:04 &  33.70 &  -18.05 & $83.9 \pm 0.66$ & 168775 & 0.574 & 0.152 & 176670 & 0.330 & 0.167
\\ 2005/06/14
   & 07:58 &  33.98 &   -1.26 & $85.0 \pm 0.90$ & 176670 & 0.521 & 0.166 & 176670 & 0.400 & 0.166
\\ 2005/06/15
   & 06:03 &  33.83 &   14.15 & $84.4 \pm 1.16$ & 176670 & 0.458 & 0.167 & 176670 & 0.457 & 0.167
\\ & 06:39 &  33.92 &    9.18 & $86.5 \pm 1.35$ & 176670 & 0.363 & 0.167 & 176670 & 0.575 & 0.166
\\ & 07:07 &  33.96 &    5.37 & $84.2 \pm 1.41$ & 176670 & 0.544 & 0.166 & 163770 & 0.392 & 0.272
\\ \hline
\end{tabular}
\end{table*}

Since the mid-1980s, great attention has been paid to Vega and other Vega-like stars. They have
been imaged from the millimetric domain down to the visible, revealing circumstellar dust arranged
in various shapes. For instance, Vega is known to be surrounded by a smooth annular structure
similar to the solar Kuiper Belt, containing about $3\times10^{-3}{\cal M}_{\oplus}$ of dust grains
\citep{Holland98,Su05}, which also shows some clumpy components \citep{Koerner01,Wilner02}.
However, due to the limitation in angular resolution of current telescopes, very little is known
about the innermost part of these debris disks, which could potentially harbour warm dust ($\gtrsim
300$~K) heated by the star as suggested by \citet{Fajardo98}. Such warm dust would have a signature
in the near- and mid-infrared that only photometric studies have attempted to detect until
recently. Indeed, Vega's near-infrared ($K$, $L$, $M$) flux was shown to be significantly above the
modelled photospheric level \citep{Mountain85}, but this discrepancy was most likely due to an
inadequate photospheric model since Vega's flux is consistent with other A-type stars to within
standard photometric precision of $2-5$\% \citep{Leggett86}. In the N band, the best constraint on
the thermal emission from warm dust has been obtained by nulling interferometry, with no resolved
emission above 2.1\% of the level of stellar photospheric emission at separations larger than
0.8~AU \citep{Liu04}. At longer wavelengths, the recent measurements obtained with Spitzer in the
far-infrared \citep{Su05} have not allowed for an investigation of the inner part of Vega's disk
because of the limited resolution (47~AU at the distance of Vega) and because hot dust is not
expected to contribute significantly to the far-infrared flux.

In this paper, we use infrared stellar interferometry to investigate the inner part of Vega's
debris disk. Such an attempt had already been made by \citet{Ciardi01}, who observed Vega with the
PTI interferometer on a 110~m long baseline in dispersed mode. The poor spatial frequency coverage
of their observations did not allow clear conclusions, although a simple model of a star and a
uniform dust disk with a $3-6$\% flux ratio was proposed to explain the observations. A more
thorough study of Vega-type stars was performed with the VLTI by \citet{diFolco04}, using short and
long baselines to separately resolve the two components of the system (stellar photosphere at long
baselines and circumstellar emission at short baselines). Unfortunately, the visibility precision
and the available baselines at the VLTI only allowed upper limits to be inferred on the flux of the
inner disks. In order to better constrain the near-infrared brightness of Vega's disk, we have used
the same method at the CHARA Array \citep{tenBrummelaar05} with an optimised set of baselines.


\section{Observations and data reduction}

Interferometric observations were obtained in the infrared $K$ band ($1.94 - 2.34$~$\mu$m) with
FLUOR, the Fiber Linked Unit for Optical Recombination \citep{Coude03}, using the S1--S2 and E2--W2
baselines of the CHARA Array, 34 and 156~metres respectively. Observations took place during Spring
2005, on May 21st and May 22nd for E2--W2, and between June 13th and June 15th for S1--S2 (see
Table~\ref{tab:obs}). The FLUOR field-of-view, limited by the use of single-mode fibers, has a
Gaussian shape resulting from the overlap integral of the turbulent stellar image with the
fundamental mode of the fiber \citep{Guyon02}. Under typical seeing conditions, it has a radius of
$1\arcsec$ (distance at which the coupling efficiency falls to 3\% of its on-axis value).

The FLUOR Data Reduction Software \citep{Coude97,Kervella04} was used to extract the squared
modulus of the coherence factor between the two independent apertures. The interferometric transfer
function of the instrument was estimated by observing calibrators before and after each Vega data
point. All calibrator stars (Table~\ref{tab:calib}) were chosen from two catalogues developed for
this specific purpose \citep{Borde02,Merand05}. Calibrators chosen in this study are all K giants,
whereas Vega is an A0 dwarf. The spectral type difference is properly taken into account in the
Data Reduction Software, even though it has no significant influence on the final result. The
efficiency of CHARA/FLUOR was consistent between all calibrators and stable night after night to
around 85\%. Data that share a calibrator are affected by a common systematic error due to the
uncertainty of the a priori angular diameter of this calibrator. In order to interpret our data
properly, we used a specific formalism \citep{Perrin03} tailored to propagate these correlations
into the model fitting process. All diameters are derived from the visibility data points using a
full model of the FLUOR instrument including the spectral bandwidth effects \citep{Kervella03}.

\begin{table}[t]
\caption{Calibrators with spectral type, $K$ magnitude, limb-darkened disk (LD) angular diameter in
$K$ band (in milliarcsec) and baseline \citep{Borde02,Merand05}.} \label{tab:calib} \centering
\begin{tabular}{ccccc}
\hline \hline               & S. type & $K$ mag & LD diam.\ (mas) & Baseline
\\ \hline \object{HD~159501} & K1 III   & 3.14   & $1.200 \pm 0.014$ & E2--W2
\\        \object{HD~163770} & K1 IIa   & 1.03   & $3.150 \pm 0.034$ & S1--S2
\\        \object{HD~165683} & K0 III   & 2.9    & $1.152 \pm 0.014$ & E2--W2
\\        \object{HD~168775} & K2 IIIab & 1.74   & $2.280 \pm 0.025$ & S1--S2
\\        \object{HD~173780} & K2 III   & 2.0    & $1.950 \pm 0.021$ & E2--W2
\\        \object{HD~176527} & K2 III   & 2.04   & $1.765 \pm 0.024$ & E2--W2
\\        \object{HD~176670} & K2.5 III & 1.6    & $2.410 \pm 0.026$ & S1--S2
\\ \hline
\end{tabular}
\end{table}


\section{Data analysis}

    \subsection{Stellar diameter}

\begin{figure}[t]
\centering \resizebox{\hsize}{!}{\includegraphics{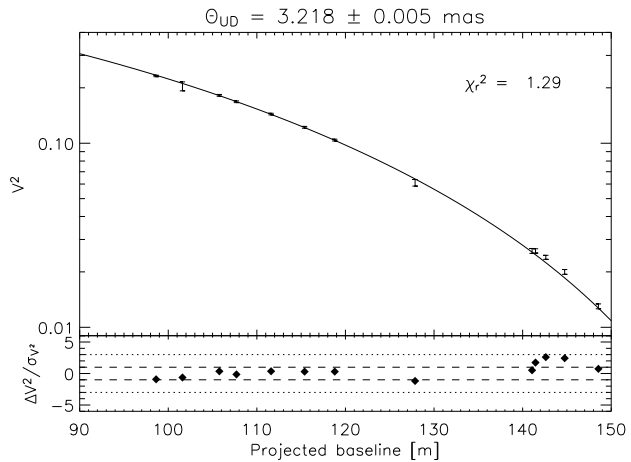}} \caption{Fit of a uniform stellar
disk model to the E2--W2 data. The quality of the fit is satisfactory (reduced $\chi^2$ of 1.29),
with small residuals that do not display any obvious trend except for a small underestimation of
the actual data for baselines between 140 and 150~m.} \label{fig:starfit}
\end{figure}

The measurements obtained with the long E2--W2 baseline are particularly appropriate for a precise
diameter determination, because they provide good spatial frequency coverage of the end of the
first lobe of the visibility curve (see Fig.~\ref{fig:starfit}). Previous interferometric
measurements obtained in the visible by \citet{Hanbury74} and \citet{Mozurkewich03} were used to
derive uniform disk (UD) diameters $\theta_{\rm UD} = 3.08 \pm 0.07$ ($\lambda=440$~nm) and
$\theta_{\rm UD} = 3.15 \pm 0.03$ ($\lambda=800$~nm) respectively. In the $K$ band, where the
limb-darkening effect is not as strong, \citet{Ciardi01} estimated the UD diameter to be
$\theta_{\rm UD} = 3.24 \pm 0.01$~mas. We have fitted a uniform stellar disk model to our E2--W2
data, assuming that Vega's photospheric intensity $I(\phi,\lambda)$ equals the Planck function with
an effective temperature of 9550~K for all angles $\phi$. The best-fit diameter is $\theta_{\rm UD}
= 3.218 \pm 0.005$~mas for an effective wavelength of 2.118~$\mu$m, which significantly revises the
previously obtained estimates\footnote{The $K$-band diameter proposed by \citet{Ciardi01} was
computed with the assumption of a flat spectrum for the Vega intensity. This explains a large part
of the discrepancy with our new value.}. The quality of the fit is quite good ($\chi^2_r=1.29$).
Unlike in the PTI data of \citet{Ciardi01}, we do not see any obvious trend in the residuals of the
fit, except for three points at projected baselines between 140 and 150~m which are slightly above
the fit (by $\sim$1.5$\sigma$). In fact, Fig.~\ref{fig:ciardi} not only shows a significant
discrepancy between the CHARA/FLUOR and the PTI data, but also between the 1999 and 2000 PTI data.
Our observations do not support the scenario of \citet{Ciardi01}, who proposed a uniform dust ring
with a $3-6$\% integrated flux relative to the Vega photosphere in $K$ band to account for the
trend that they observed in the residuals of the fit obtained with a simple limb-darkened disk (LD)
stellar model.

Note that fitting an LD stellar model to our data would only marginally improve the fit (see
Table~\ref{tab:ld}), as the shape of the first-lobe visibility curve is not very sensitive to limb
darkening. Moreover, the actual limb-darkening parameter may be significantly larger than standard
tabulated values because Vega is suspected to be a fast rotating star viewed nearly pole-on and the
equatorial darkening may bias the limb profile \citep{Gulliver94,Peterson04}. Complementary
observations to our data set, obtained by \citet{Aufdenberg06} at $\sim$250~m baselines, confirm
this fact and lead to an accurate estimation of the $K$-band limb profile, which mostly affects
visibilities beyond the first null and will not be discussed here.

    \subsection{Visibility deficit at short baselines} \label{sub:deficit}

\begin{figure}[t]
\centering \resizebox{\hsize}{!}{\includegraphics{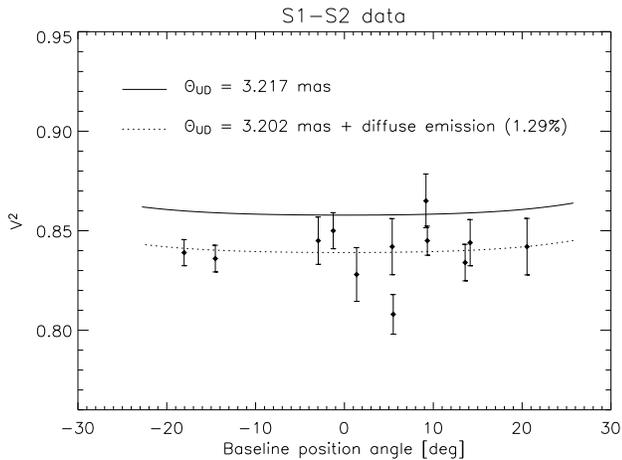}} \caption{The data obtained with
the S1--S2 baseline ($\sim$34~m) are displayed as a function of the projected baseline's position
angle together with the best UD fit computed over the whole data set (3.217~mas). The data points
are significantly below the best UD fit, with a mean visibility deficit $\Delta V^2 \simeq 2$\%.
The addition of a uniform diffuse source of emission in the FLUOR field-of-view reconciles the best
fit with the data (dotted line). Note that there is no obvious dependence of the data points with
respect to position angle, which would be indicative of an asymmetric extended emission.}
\label{fig:s1s2}
\end{figure}

With this precise diameter estimation, we can now have a look at the short-baseline data. In fact,
these points do not significantly contribute to the UD fit because of the low spatial frequencies
they sample. Including all the data points in the fitting procedure gives a best-fit diameter
$\theta_{\rm UD} = 3.217 \pm 0.013$~mas, but with a poor $\chi^2_r=3.36$. We show the reason for
this poor reduced $\chi^2$ in Fig.~\ref{fig:s1s2}, where the S1--S2 data points are plotted as a
function of position angle together with the best UD fit (solid line). The observations are
consistently below the fit, with a $\Delta V^2 = 1.88 \pm 0.34 $\%.

Systematic errors in the estimation of the calibrator diameters or limb-darkened profiles are
possible sources of bias in interferometric observations. In order to explain the measured
visibility deficit in the S1--S2 data, the diameters of the three short-baseline calibrators
(Table~\ref{tab:calib}) should have been underestimated by 0.26, 0.35 and 0.33~mas respectively,
which represent about 10 times the estimated error on their diameters. We have made sure that such
improbable errors were not present in our calibration procedure by cross-calibrating the three
calibrators. No significant departure from the expected LD diameters was measured, and the
calibrated visibilities of Vega do not depend on the chosen calibrator. Therefore, it appears
extremely unlikely that the calibration process may have induced the observed visibility deficit.

\begin{figure}[t]
\centering \resizebox{\hsize}{!}{\includegraphics{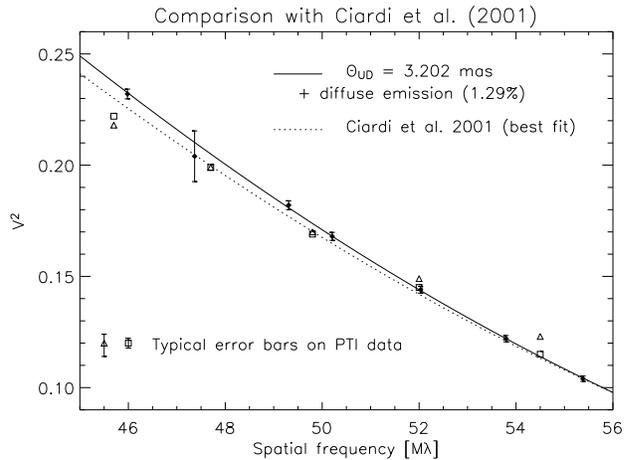}} \caption{Comparison of our E2-W2
data (black dots) with the observations of \citet{Ciardi01} obtained at PTI (triangles: data
acquired in 1999, squares: data acquired in 2000). The data are displayed as a function of spatial
frequency, taking an equivalent wavelength of 2.145~$\mu$m for the FLUOR instrument (computed for a
flat stellar spectrum as in the study of \citet{Ciardi01}). The 1$\sigma$ errors on the PTI data
are shown at the bottom of the figure for the sake of clarity.} \label{fig:ciardi}
\end{figure}

A limb-darkened stellar model for Vega will not reconcile the best-fit stellar model with the
S1--S2 data points (see Table~\ref{tab:ld}), because low spatial frequencies are not sensitive to
limb darkening. One may think of stellar asymmetry as a possible reason for the visibility deficit
at short baselines, since the position angles of the short and long baselines are almost
perpendicular (see Table~\ref{tab:obs}). However, an oblateness ratio of 1.07 for Vega would be
needed to explain the deficit, which would strongly contradict previous interferometric studies
\citep{vanBelle01,Peterson04}. Other stellar features such as spots would not explain this deficit
either as they can only appear in the second and higher lobes of the visibility function. In fact,
a natural explanation to the observed visibility deficit would be the presence of an extended
source of emission in the interferometric field-of-view (e.g.\ disk or companion), which would be
resolved with the S1--S2 baseline (i.e., incoherent emission).

In order to assess the amount of incoherent emission needed to explain the observed visibility
deficit, we have added a diffuse emission, uniformly distributed in the CHARA/FLUOR field-of-view,
to our UD stellar model. Fitting this new model to the complete data set gives the following final
result: $\theta_{\rm UD} = 3.202 \pm 0.005$~mas, $K$-band flux ratio = $1.29 \pm 0.19$\%, with a
significantly decreased $\chi^2_r = 1.10$ (instead of 3.36). This result is almost independent of
the extended source morphology, as the spatial frequency coverage of our interferometric data is
too scarce to constrain its spatial distribution. The extended structure, detected with very good
confidence (almost 7$\sigma$), would thus have a relative flux contribution of 1.29\% with respect
to the Vega photosphere in $K$ band when integrated over the whole field-of-view (7.8~AU in
radius). Such an excess does not contradict photometric measurements in the $K$ band, which have
typical accuracies of $2-3$\% \citep{Megessier95}. The result of the fit is displayed in
Fig.~\ref{fig:s1s2} (dotted line) and Fig.~\ref{fig:ciardi} (solid line), as well as in
Fig.~\ref{fig:disk} for a realistic debris disk model (see Sect.~\ref{sub:disk}), which gives the
same best-fit parameters.

\begin{table}[t]
\caption{Influence of the limb-darkening parameter $\alpha$ on the best-fit diameter and the
associated reduced $\chi^2$ using the whole data set, assuming a brightness distribution $I (\mu) =
\mu^{\alpha}$ with $\mu = \cos \theta$ the cosine of the azimuth of a surface element of the star
\citep{Hestroffer97}. The visibility deficit measured at short baselines (S1--S2) with respect to
the best-fit model is given in the last column, showing a weak dependence on the limb-darkening
model.} \label{tab:ld} \centering
\begin{tabular}{cccc}
\hline \hline            & Best-fit                & $\chi^2_r$ & $\Delta V^2$
\\              $\alpha$ & $\theta_{\rm LD}$ (mas) & (all data) & (S1--S2)
\\ \hline          0.0   & $3.217 \pm 0.013$       &    3.36    & 1.88\%
\\                 0.1   & $3.264 \pm 0.013$       &    3.14    & 1.83\%
\\                 0.2   & $3.310 \pm 0.012$       &    2.96    & 1.78\%
\\                 0.3   & $3.356 \pm 0.012$       &    2.82    & 1.73\%
\\                 0.4   & $3.402 \pm 0.011$       &    2.71    & 1.67\%
\\                 0.5   & $3.447 \pm 0.011$       &    2.64    & 1.62\%
\\                 0.6   & $3.491 \pm 0.011$       &    2.61    & 1.58\%
\\                 0.7   & $3.536 \pm 0.012$       &    2.60    & 1.53\%
\\                 0.8   & $3.579 \pm 0.012$       &    2.62    & 1.49\%
\\                 0.9   & $3.623 \pm 0.012$       &    2.66    & 1.44\%
\\ \hline
\end{tabular}
\end{table}


\section{Discussion}

In this section, we discuss the possible sources of incoherent flux around Vega that could account
for the observed visibility deficit at short baselines.

  \subsection{Point source}

Because of our sparse sampling of spatial frequencies, a point source located in the FLUOR
field-of-view could also be the origin of the observed visibility deficit. Regardless of the bound
or unbound character of the companion, there are essentially two regimes to be considered when
computing the visibility of a binary system, depending on whether the fringe packet associated with
the companion falls into the FLUOR observation window or not. The observation window is defined as
the total optical path $L_{\rm OPD}$ scanned by the FLUOR dither mirror, which is used to
temporally record the fringes. The secondary fringe packet lies outside the observation window if
$|B \, \alpha \cos\theta| > L_{\rm OPD}/2$, where $B$ is the baseline length, $\alpha$ the angular
separation of the binary system, $\theta$ the angle between the baseline and the orientation of the
binary system, and $L_{\rm OPD}=102$~$\mu$m. In that case, e.g.\ for an angular separation larger
than 350~mas at a baseline of 34~m, the flux from the secondary will contribute incoherently and
will lead to the same signature as a diffuse emission in the FLUOR field-of-view. A binary star
with a separation ranging between 350 and 1000~mas could therefore reproduce the observed
visibilities. On the other hand, if the secondary fringe packet is inside the observation window,
it will lead either to a visibility modulation of twice the flux ratio as a function of baseline
azimuth if the fringe packets are superposed, or to an enhancement of the measured visibility if
the fringe packets are separated. Even if such behaviour does not seem compatible with the observed
visibilities, our sparse data cannot definitely rule out a solution with a close companion.

The presence of a point source located within the FLUOR field-of-view could thus possibly explain
our observations. The minimum $K$-band flux ratio between the point source and Vega is $1.29 \pm
0.19$\%, valid for a very close companion ($\lesssim 50$~mas). Because of the Gaussian shape of the
off-axis transmission, the companion would have a larger flux if located farther away from the
star. For instance, the flux should be increased by 10\% at 100~mas, by 50\% at 200~mas and by
3000\% at 500~mas from the star in order to reproduce the observed visibility deficit. Based on a
minimum $K$-band flux ratio of $1.29 \pm 0.19$\% and a $K$ magnitude of 0.02 for Vega
\citep{Megessier95}, we deduce an upper limit of $K=4.74 \pm 0.17$ for a companion.

    \subsubsection{Field star}

Although Vega is known to be surrounded by a number of faint objects ($V>9$) with low proper motion
since the beginning of the 20th century \citep{Dommanget02}, these objects are far enough from Vega
(at least $1\arcmin$) so that they do not interfere with our measurements. In the infrared, neither
adaptive optics studies \citep{Macintosh03,Metchev03} nor the 2MASS survey \citep{Cutri03}
identified any $K<5$ object within $1\arcmin$ of Vega. In fact, the local density of such objects
is about $5 \times 10^{-4}$ per arcmin$^2$ according to the 2MASS survey, so that the probability
to find a $K<5$ source within $1\arcsec$ of Vega is smaller than $4.3 \times 10^{-7}$.

    \subsubsection{Physical companion}

At the distance of Vega, the putative companion would have a maximum absolute magnitude $M_K=5.15
\pm 0.17$. Assuming this companion to be a star of the same age as Vega itself, comprised between
267 and 383~Myr \citep{Song01}, we use the evolutionary models developed by \citet{Baraffe98} to
deduce the range of effective temperature and mass for the companion: $T_{\rm eff} = 3890 \pm 70$~K
and ${\cal M} = 0.60 \pm 0.025$~${\cal M}_{\odot}$. This roughly corresponds to an M0V star
\citep{Delfosse00}.

With a $V-K$ of 3.65 \citep{Bessell88}, the M0V companion would have a $V$ magnitude of 8.41 and
would therefore have remained undetected in high resolution visible spectra of Vega (M.~Gerbaldi,
personal communication). Adaptive optics studies in the near-infrared would not have noticed the
companion either, due to its very small angular distance from the bright Vega ($<1\arcsec$). At
longer wavelengths, the expected infrared excess due to an M0V companion is not large enough to be
detected by classical photometry as its does not exceed 2\% between 10 and 100~$\mu$m. Indirect
methods are in fact much more appropriate to detect this kind of companion.

Astrometric measurements of Vega with Hipparcos did not detect the presence of any companion, with
an astrometric precision of 0.5~mas \citep{Perryman97}. With a mass ratio of 4.2 between Vega ($2.5
{\cal M}_{\odot}$) and its putative M0V companion ($0.6 {\cal M}_{\odot}$), a $3\sigma$ astrometric
stability of 1.5~mas implies that the orbital semi-major axis of the putative companion cannot be
larger than 6.3~mas\footnote{The astrometric signature of a low-mass companion is given by the
ratio between the orbital semi-major axis and the mass ratio \citep{Perryman00}.} ($=0.05$~AU $= 4
R_{\star}$) with a 99\% confidence assuming a circular orbit, which is anticipated for such a small
separation. Such a close companion, which could also fit the interferometric data, would have an
appreciable signature in radial velocity measurements, unless the binary system is seen almost
exactly pole-on. Precise measurements recently obtained with the ELODIE spectrometer have shown a
relative stability of Vega's radial velocity over several months, with amplitudes lower than
100~m/s and a precision of order of 30~m/s each (F.~Galland, private communication). Assuming that
the orbital plane of the M0V companion is perpendicular to Vega's rotation axis, inclined by
$5.1\degr$ with respect to the line-of-sight \citep{Gulliver94}, the companion should be farther
than 80~AU from Vega to be compatible with the measured radial velocity stability. In fact, for an
M0V companion at 0.05~AU from Vega not to display any radial velocity signature at the 100~m/s
level, its orbital inclination needs to coincide with the plane of the sky to within $\pm
0.13\degr$ \citep{Perryman00}. Even if such an inclination is possible, the probability for the
system to be so close to pole-on is very low (it ranges between about $6 \times 10^{-4}$ and
$10^{-6}$ depending on the assumptions on the statistical distribution of low-mass companion
orbital planes). In conclusion, even though the presence of an M0V companion close to Vega could
explain the interferometric data, there is strong evidence that such a companion does not exist.

  \subsection{Circumstellar material} \label{sub:disk}

Circumstellar disks around MS stars are understood to be composed of second-generation dust grains
originating from collisions between small bodies (asteroids) or from the evaporation of comets
\citep{Backman93}. They are assumed to be continuously replenished since dust grains have a limited
lifetime ($<10$~Myr) due to radiation pressure, Poynting-Robertson (P-R) drag and collisional
destruction \citep{Dominik03}. Several studies have shown Vega to harbour a cold circumstellar dust
ring $\sim$85~AU in radius \citep{Holland98,Heinrichsen98,Koerner01,Wilner02}. \citet{Su05}
interpreted the extended dust emission (up to 600~AU, i.e., $77\arcsec$) detected by Spitzer as the
signature of dust grains being expelled by radiation pressure from the Vega system as a result of a
recent collision in the main planetesimal ring and subsequent collisional cascade. Even if the
presence of dust in the inner part of the disk has not been detected yet due to instrumental
limitations, an equivalent to the solar zodiacal cloud is expected to be found around Vega. The
thermal and scattered emissions from warm grains surrounding Vega could thus be a natural
explanation to the visibility deficit observed at short baselines, provided that a sufficient
quantity of dust is present within 8~AU from the star.

\begin{figure}[t]
\centering \resizebox{\hsize}{!}{\includegraphics{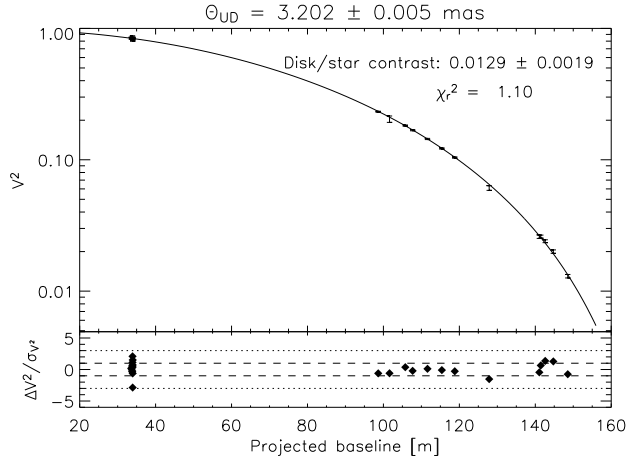}} \caption{Fit of a uniform stellar
disk + circumstellar disk model to our full data set, using the model of \citet{Kelsall98}.}
\label{fig:disk}
\end{figure}

In order to assess the adequacy of a circumstellar disk to reproduce the observations, we have
fitted our full data set with the only known model for an inner debris disk, i.e., the zodiacal
disk model of \citet{Kelsall98}\footnote{This model was implemented in an IDL package called
ZODIPIC by M.~Kuchner (http://www.astro.princeton.edu/$\sim$mkuchner/).}, assuming that the inner
dust distribution around Vega follows the same density and temperature power-laws as for the solar
zodiacal cloud. The result is displayed in Fig.~\ref{fig:disk} wherein all data points are nicely
spread around the best-fit model (as expected, because our interferometric data are not sensitive
to the particular morphology of the incoherent emission). The long-baseline data are also better
fitted than with a simple UD model, because the presence of the dust disk has some influence on the
slope of the visibility curve at long baselines \citep{diFolco04}. The resulting flux ratio between
the whole circumstellar disk and the stellar photosphere ($1.29 \pm 0.19$\%) is the same as with a
simple model of uniform diffuse emission (Sect.~\ref{sub:deficit}), with the same reduced $\chi^2$
of 1.10. Using the model of \citet{Kelsall98}, a flux ratio of 1.29\% in $K$ band would suggest
that the dust density level in the inner Vega system is about 3000 times larger than in the solar
zodiacal cloud. However, we will see later on that this model is not appropriate to represent
Vega's inner disk (it would largely overestimate its mid-infrared flux), so that the comparison is
not actually pertinent.

        \subsubsection{Physical properties of the dust grains}

Let us now try to evaluate the main physical properties of the dust grains in the inner debris
disk. Table~\ref{tab:photo} gives the photometric constraints on the near- and mid-infrared excess
flux around Vega currently available in the literature. Photometric constraints at wavelengths
longer than 12~$\mu$m are not appropriate for our purpose as they are mostly sensitive to the cold
outer disk (the inner disk is not supposed to produce a significant photometric contribution in the
far-infrared). The large error bars on the photometric measurements take into account both the
actual error on photometric measurements and the estimated accuracy of photospheric models for
Vega, to which the measurements are compared. Our study is compatible with previous near-infrared
measurements but provides a much stronger constraint on the inner disk, because interferometry
spatially resolves the disk from the stellar photosphere and focuses on the inner part of the disk
thanks to the small field-of-view. Nulling interferometry at the MMT with the BLINC instrument also
provides a valuable constraint on the mid-infrared excess \citep{Liu04}. The sinusoidal
transmission map of this nulling interferometer restricts however the observation to the part of
the disk located farther than about 125~mas ($\sim$~1~AU) from the star. This explains why the
result of this study is significantly below the estimated mid-infrared photometric excesses, as it
is not sensitive to hot grains in the innermost part of the disk.

\begin{table}[t]
\caption{Available constraints on the near- and mid-infrared excess around Vega. References: (1)
\citet{Campins85}; (2) \citet{Blackwell83}; (3) \citet{Rieke85}; (4) \citet{Liu04}; (5)
\citet{Cohen92}, with the absolute photometric error estimated by \citet{Aumann84}. The photometric
data in references (1), (2) and (3) have been compared to the most recent Kurucz photospheric model
of Vega \citep{Bohlin04}, which has a typical uncertainty of 2\% in the infrared (this uncertainty
has been added to the estimated errors on the measurements). Note that the interferometric data
from FLUOR and BLINC only sample a specific part of the inner disk, while the photometric studies
include Vega's entire environment.} \label{tab:photo} \centering
\begin{tabular}{cccl}
\hline \hline  Wavelength  & Excess            & Instruments & References
\\ \hline      1.26 $\mu$m & $2.4 \pm 2.9$\%   & Catalina, UKIRT  & (1), (2)
\\             1.60 $\mu$m & $-2.4 \pm 3.6$\%  & Catalina    & (1)
\\             2.12 $\mu$m & $1.29 \pm 0.19$\% & CHARA/FLUOR       & This study
\\             2.20 $\mu$m & $4.1 \pm 3.0$\%   & Catalina, UKIRT  & (1), (2)
\\             3.54 $\mu$m & $3.1 \pm 3.0$\%   & Catalina, UKIRT  & (1), (2)
\\             4.80 $\mu$m & $7.1 \pm 5.1$\%   & Catalina, UKIRT  & (1), (2)
\\             10 $\mu$m   & $6 \pm 4.5$\%     & Various     & (3)
\\             10.6 $\mu$m & $0.2 \pm 0.7$\%   & MMT/BLINC       & (4)
\\             12 $\mu$m   & $1.2 \pm 5$\%     & IRAS       & (5)
\\ \hline
\end{tabular}
\end{table}

\begin{figure}[t]
\centering \resizebox{\hsize}{!}{\includegraphics{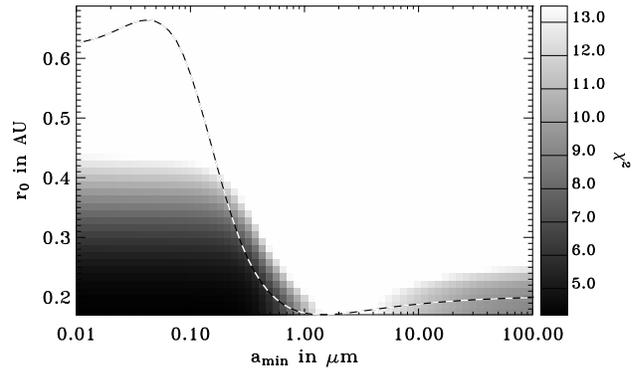}} \caption{Map of the $\chi^2$ as a
function of minimum grain size $a_{\rm min}$ and inner radius $r_0$, obtained by fitting the
circumstellar disk model of \citet{Augereau99} to the SED data of Table~\ref{tab:photo}. We have
assumed a surface density power-law $\Sigma(r) \propto r^{-4}$ and a grain size distribution $dn(a)
\propto a^{-3.7} da$, with a maximum size of 1500~$\mu$m. In this simulation, the disk is composed
of 50\% amorphous carbons and 50\% glassy olivines (see text). The dashed line represents the
distance at which sublimation happens for dust grains of a given size (isotherm $T=1700$~K). The
axis labels should therefore read ``$a$ in $\mu$m'' and ``sublimation radius in AU'' for this
curve.} \label{fig:chi2}
\end{figure}

We have tried to reproduce the Spectral Energy Distribution (SED) of the infrared excess as listed
in Table~\ref{tab:photo} with the debris disk model developed by \citet{Augereau99}. For that
purpose, we took for Vega a NextGen model atmosphere spectrum \citep{Hauschildt99} with $T_{\rm
eff}=9600$~K and $\log(g)=4.0$, scaled to match the observed visible magnitude ($V=0.03$) at a
distance of 7.76~pc, which gives a luminosity of $58.7 L_{\odot}$. Various grain compositions and
size distributions were used in the disk model, as well as various radial density profiles,
assuming no azimuthal dependence. In each model, the sublimation temperature of the grains is set
to $T_{\rm sub} = 1700$~K. At a given distance and for a given size distribution, only grains large
enough to survive the sublimation process can actually coexist (see dashed curve in
Fig~\ref{fig:chi2}). The normalised differential size distribution between $a_{\rm min}$ and
$a_{\rm max}$ (fixed) is thus truncated at $a_{\rm sub}$, which depends on the radial distance to
the star. For each model, a $\chi^2$ map is computed for all possible values of $a_{\rm min}$
(minimum grain size) and $r_0$ (inner radius where the disk is artificially truncated), adjusting
the surface density at $r_0$ by a least-squares method (see Fig.~\ref{fig:chi2}). The most
constraining observations in this process are the two interferometric measurements at 2.12 and
10.6~$\mu$m, so that the fitting procedure mainly boils down to adjusting the near-infrared flux
without producing a too strong 10.6~$\mu$m emission feature. Comparison of $\chi^2$ values allowed
us to infer most probable physical properties for the inner debris disk.

\begin{itemize}
\item {\bf Size distribution:} The inner disk seems to be mainly composed of hot ($\sim$1500~K) and
small ($<1$~$\mu$m) dust grains, which emit mostly in the near-infrared. Although larger grains
($\ge 10$~$\mu$m) cannot be ruled out as the main source of the excess, such grains generally
produce too large a mid-infrared flux as they emit more efficiently in this wavelength range. This
suggests a steep size distribution with a small minimum grain size ($a_{\rm min} \leq 0.3$~$\mu$m,
assuming compact grains). For instance, we find that a size distribution similar to that inferred
by \citet{Hanner84} for cometary grains provides a good fit to the SED, as well as the interstellar
size distribution of \citet{Mathis77}. Both have power-law exponents of $-3.5$ or steeper. On the
contrary, the size distribution of \citet{Grun85} for interplanetary dust particles does not
provide a good reproduction of the disk's SED, so that the grain size distribution is most probably
different from that of the solar zodiacal cloud described by \citet{Reach03}.
\item {\bf Composition:} Large amounts of highly refractive grains, such as graphites
\citep{Laor93} or amorphous carbons \citep{Zubko96}, are most probably present in the inner disk.
This is required in order to explain the lack of significant silicate emission features around
10~$\mu$m \citep{Gaidos04}, which are especially prominent for small grains. Silicate grains can
still be present in the disk, but with a maximum volume ratio of $\sim$70\%, using the astronomical
silicates of \citet{Weingartner01} or the glassy olivines (Mg$_{2y}$Fe$_{2-2y}$SiO$_4$) of
\citet{Dorschner95} with $y=0.5$. This is another difference from the solar zodiacal cloud, which
is thought to contain about 90\% of silicate grains \citep{Reach03}. Such a mixing ratio would only
be possible around Vega if the grains were sufficiently big ($a_{\rm min} \ge 10$~$\mu$m), so that
the silicate emission feature around 10~$\mu$m would not be too prominent.
\item {\bf Density profile:} The inner radius $r_0$ of the dusty disk is estimated to be
between 0.17 and 0.3~AU. Assuming a sublimation temperature of 1700~K, dust grains larger than
0.5~$\mu$m would survive at such distances (see dashed curve in Fig.~\ref{fig:chi2}) while smaller
grains, which are hotter, sublimate farther from the star (e.g.\ at $\sim$0.6~AU for a 0.1~$\mu$m
grain). A steep power-law for the radial surface density distribution has also been inferred from
our investigations. A power-law exponent of $-4$ or steeper provides a good fit to the SED, as it
reduces the amount of dust in the regions farther than 1~AU and thereby explains the non-detection
with MMT/BLINC reported by \citet{Liu04}. In contrast, the zodiacal disk model of \citet{Kelsall98}
has a flat surface density power-law with an exponent around $-0.34$.
\end{itemize}

\begin{figure}[t]
\centering \resizebox{\hsize}{!}{\includegraphics{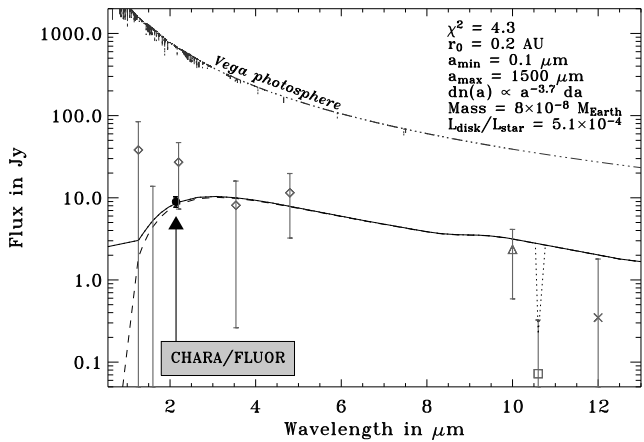}} \caption{A possible fit of our
debris disk model \citep{Augereau99} to the photometric and interferometric constraints of
Table~\ref{tab:photo}: the diamonds correspond to references (1) and (2), the filled circle to this
study, the triangle to (3), the square to (4) and the cross to (5). The model used here has a size
distribution $dn(a) \propto a^{-3.7} da$ with limiting grain sizes $a_{\rm min} = 0.1$~$\mu$m and
$a_{\rm max} = 1500$~$\mu$m, a surface density power-law $\Sigma(r) \propto r^{-4}$ with an inner
radius $r_0 = 0.2$~AU, and assumes a disk composed of 50\% amorphous carbon and 50\% glassy
olivine. The solid and dotted lines represent the total emission from the disk on a 8~AU
field-of-view, respectively without and with the spatial filtering of interferometric studies,
while the dashed line takes only the thermal emission into account. The photospheric SED, simulated
by a NextGen model atmosphere (see text), is represented as a dashed-dotted line for comparison.}
\label{fig:SED}
\end{figure}

Using these most probable parameters for the inner disk and a mixed composition of 50\% amorphous
carbons \citep{Zubko96} and 50\% glassy olivines \citep[MgFeSiO$_4$, ][]{Dorschner95}, we have
obtained a relatively good fit to the SED as illustrated in Fig.~\ref{fig:SED}, where we see that
the thermal emission from the hot grains supersedes the contribution from scattered light at
wavelengths longer than 1.3~$\mu$m. Based on our model and assuming a size distribution $dn(a)
\propto a^{-3.7} da$ with $a_{\rm min} = 0.1$~$\mu$m and $a_{\rm max}=1500$~$\mu$m, we can deduce
estimations for the dust mass in the inner 10~AU of the disk (${\cal M}_{\rm dust} \sim 8 \times
10^{-8} {\cal M}_{\oplus}$, equivalent to the mass of an asteroid about 70~km in diameter) and for
the bolometric luminosity ratio between the inner disk and the star ($L_{\rm disk} / L_{\star} \sim
5 \times 10^{-4}$). Because of the high temperature of the grains, the luminosity of the inner disk
is more than one order of magnitude larger than the luminosity of the outer disk estimated by
\citet{Heinrichsen98}, even though it is almost $10^5$ times less massive than the outer disk.
These results need to be confirmed by future studies, as the SED of the inner disk is still
relatively poorly constrained. They have been included in this paper to demonstrate that the
presence of warm circumstellar dust can reproduce the various observations, and to provide a
plausible dust-production scenario as discussed below.

        \subsubsection{A possible scenario for the presence of hot dust}

In fact, three main scenarios may explain the presence of small dust grains so close to Vega. As in
the case of the solar zodiacal cloud, they could be produced locally, e.g.\ by collisions between
larger bodies arranged in a structure similar to the solar asteroidal belt. Another local source of
small grains is the evaporation of comets originating from the reservoir of small bodies at
$\sim$85~AU from Vega or from an inner population of icy bodies as in the case of $\beta$~Pic
\citep{Beust00}. Finally, dust grains produced by collisions in the outer disk could drift towards
the inner region because of P-R drag. However, this latter scenario cannot be connected to the
recent collision(s) in the outer disk suggested by \citet{Su05}, because of the long timescale of
P-R drag \citep[$2 \times 10^7$~yr, ][]{Dent00}. Moreover, due to the much shorter collisional
timescale ($5 \times 10^5$~yr in the outer disk), this process is not very efficient and is
therefore unlikely to produce the observed amount of dust in Vega's inner system. Our observations
cannot discriminate between the two remaining scenarios, even though a cometary origin is favoured
by the steep size distribution of dust grains \citep{Hanner84} and by the small inner disk radius.

Due to radiation pressure, small grains will not survive in the Vega inner disk more than a few
years before being ejected toward cooler regions \citep{Krivov00}. Larger grains would survive
somewhat longer, but not more than a few tens of years due to the high collision rate in the inner
disk. A large dust production rate ($\sim 10^{-8} {\cal M}_{\oplus}$/yr) is thus needed to explain
our observations, suggesting that major dynamical perturbations are currently ongoing in the Vega
system. An attractive scenario would be an equivalent to the Late Heavy Bombardment that happened
in the solar system in the $700$~Myr following the formation of the planets \citep{Hartmann00},
i.e., at a period compatible with the age of Vega ($\sim$350~Myr). Such a bombardment, most
probably triggered by the outward migration of giant planets \citep{Gomes05}, could explain the
presence of small grains around Vega both in its outer disk, due to an enhanced collision rate in
this part of the disk, and in its inner disk, due to the high number of comets sent toward the star
by gravitational interaction with the migrating planets. Although the presence of giant planets
around Vega has not been confirmed yet, \citet{Wyatt03} has suggested that the outward migration of
a Neptune-sized body from 40 to 65~AU could explain the observed clumpy structure in Vega's outer
disk.


\section{Conclusion}

In this paper, we have presented high precision visibility measurements obtained on Vega at the
CHARA Array with the FLUOR beam-combiner. The presence of a significant deficit of visibility at
short baselines with respect to a simple uniform disk stellar model led us to the conclusion that
an additional source of $K$-band emission is present in the FLUOR field-of-view centred around Vega
($1\arcsec$ in radius), with an estimated excess of $1.29\pm 0.19$\% relative to the photospheric
emission. Among the possible sources for this excess emission, the presence of dust grains in the
close vicinity of Vega, heated by the star and radiating mostly in the near-infrared, is proposed
as the most probable one. Vega, a prototypical debris-disk star surrounded by a large quantity of
dust at about 85~AU, was already suspected by several authors to harbour warm dust grains arranged
in an inner circumstellar disk. Previous studies were however limited to a precision of a few
percent on the total infrared flux of the Vega system and therefore did not provide a precise
estimation of the excess emission associated with the inner disk.

Thanks to our precise determination of the integrated $K$-band emission emanating from the inner
8~AU of the Vega debris disk, we were able to infer some physical properties of the dust, which is
suspected to be mainly composed of sub-micronic highly refractive grains mainly concentrated in the
first AU around Vega and heated up to 1700~K. An estimated dust mass of $8 \times 10^{-8} {\cal
M}_{\oplus}$ and a fractional luminosity of $\sim 5 \times 10^{-4}$ are derived from our best-fit
model. We propose that a major dynamical event, similar to the solar system Late Heavy Bombardment,
might be the cause for the presence of small dust grains in the inner disk of Vega.

\begin{acknowledgements}
We thank P.J.~Goldfinger and G.~Romano for their assistance with the operation of CHARA and FLUOR
respectively. The CHARA Array is operated by the Center for High Angular Resolution Astronomy with
support from Georgia State University and the National Science Foundation, the W.M.\ Keck
Foundation and the David and Lucile Packard Foundation. This research has made use of NASA's
Astrophysics Data System and of the SIMBAD database, operated at CDS (Strasbourg, France).
\end{acknowledgements}


\bibliographystyle{aa} 
\bibliography{4522} 

\end{document}